# Tilted and Non Tilted C-Field Cosmological Models

**D. D. Pawar and V. J. Dagwal*, S. P. Shahare***

School of Mathematical Sciences, Swami Ramanand Teerth
Marathwada University, Vishnupuri Nanded-431 606 (India)
* Dept. of Mathematics, Govt. College of Engineering, Amravati 444 604, India
**Pote College of Engineering and Management, Amravati-444604, India
E-mail: dypawar@yahoo.com, vdagwal@gmail.com

**Abstract:**

Tilted and non tilted homogeneous plane symmetric C-field cosmological models are investigated. Using the method of Narlikar and Padmanabhan [4], the solutions have been presented when the creation field C is function of time *t* only. To get the deterministic model, we have assumed the supplementary conditions $p = \rho$. The behaviour of different stages of the universe has been studied in non tilted models. We have also investigated the behaviours of some physical parameters.

## 1. Introduction:

The universe does not have any singular beginning or an end on the cosmic time scale in C-field (Creation field) theory. Bondi and Gold [1] investigated steady-state theory; there is no big-bang type singularity in C-field theory. Hoyle and Narlikar [2] adopted a field theoretic approach for the matter creation, by introducing a mass less and charge less scalar field. Matter creation is accomplished at the expense of negative energy C-field has obtained by Narlikar [3]. A solution of Einstein's field equations which admit relation and a negative energy massless scalar creation field as a source have examined by Narlikar and Padmanabhan [4]. Chatterjee and Banerjee [5] have investigated C-field cosmology in higher dimensional. Bali and Tikeker [6] discussed the C-field cosmology with variable G in the flat FRW model. Modelling repulsive gravity with creation have presented by Vishwakarma and Narlikar [7]. C-field cosmological models based on Hoyle-Narlikar theory with variable gravitational constant using FRW space-time for positive and negative curvature for dust distribution have studied by Bali and Kumawat [8]. Bianchi type-I, III, V,VI and Kantowski-Sachs universes in creation-field cosmology have obtained by Singh and Chaubey [9]. Adhav et al. [10] have discussed Kasner Universe in Creation-Field Cosmology.

Homogeneous and anisotropic cosmological models have been studied widely in the framework of general relativity. These models are more restricted than the inhomogeneous models. But in spite of this, they explain a number of observed phenomena quite satisfactorily. In recent years, there has been a considerable interest in investigating spatially homogeneous and anisotropic cosmological models in which matter does not move orthogonal to the hyper surface of homogeneity. Such types of models are called tilted cosmological models. King and Ellis [11]; Ellis and King [12]; Collins and Ellis [13] have presented the general dynamics of tilted cosmological models. Dunn and Tupper [14] have calculated tilted Bianchi type-I cosmological model for perfect fluid. Tilted electromagnetic Bianchi type-I cosmological model in General Relativity studied by Lorentz [15]. Bianchi type-I cosmological model with heat flux in General Relativity obtained different aspects of tilted cosmological models has examined by Mukherjee [16]. Lidsey [17], Hewitt et al. [18], Horwood et al [19] and Apostolopoulos [20] have examined different aspects of tilted cosmological models. Coley and Hervik [21] have studied Bianchi cosmologies a Tale of two tilted fluids. Pradhan and Srivastava [22] have presented tilted Bianchi type v bulk viscous cosmological models in general relativity. Pawar et al. [23] have studied tilted plane symmetric cosmological models with heat conduction and disordered radiation. Conformally flat tilted cosmological models, Kantowaski-Sachs cosmological models are studied by Pawar and Dagwal [24, 25]. Bianchi Type I tilted cosmological model for barotropic perfect fluid distribution with heat conduction in General Relativity presented by Bali and Kumawat [26]. Tilted Plane Symmetric vulk viscous cosmological model with varying $\Lambda$–Term investigated by Bhaware et al. [27]. Tilted Bianchi type $VI_0$ cosmological model in Saez and Ballester scalar tensor theory of gravitation obtained by Sahu [28]. Recently two fluids tilted cosmological model in General Relativity presented by Pawar and Dagwal [29]. Pawar et al. studied tilted plane symmetric magnetized cosmological models [30]. The stability of non-tilted Bianchi models against tilt investigated by Barrow and Hervik [31]. Non-tilted Bianchi VII0 models the radiation fluid presented by Nilsson [32].

## 2. Metric and Field Equation:

We consider the metric (Pawar and Solanke [33]) in the form

$$ds^2 = dt^2 - R^2 \left\{ dx^2 + dy^2 + \left(1 + \beta \int \frac{dt}{R^3}\right)^2 dz^2 \right\}, \qquad (1)$$

where R is functions of *t* alone.

The Einstein's field equations are

$$R_i^j - \frac{1}{2} g_i^j R = -\left( T_{i\,(m)}^j + T_{i\,(c)}^j \right). \qquad (2)$$

where $T_{i\,(m)}^j$ is the energy momentum tensor for perfect fluid and $T_{i\,(c)}^j$ is the energy momentum tensor for creation field are given by

$$T_i^j{}_{(m)} = (p+\rho)u_i u^j - p\, g_i^j + q_i u^j + u_i q^j, \tag{3}$$

$$T_i^j{}_{(c)} = -f\left(C_i C^j - \frac{1}{2} g_i^j C^k C_k\right), \tag{4}$$

with $\quad g^{ij} u_i u_j = 1,$ (5)

$$q_i q^j > 0, \quad q_i u^i = 0. \tag{6}$$

Where $p$ is the pressure, $\rho$ is the energy density, $q_i$ is the heat conduction vector orthogonal to $u_i$. The fluid vector $u_i$ has the components $(R\sinh\alpha, 0, 0, \cosh\alpha)$ satisfying Equation (5) and $\alpha$ is the tilt angle. $f > 0$ is the coupling constant between matter and creation field and $C_i = \dfrac{dC}{dx^i}$.

The Einstein's field equation (2) reduces to

$$\frac{R_4^2}{R^2} + 2\frac{R_{44}}{R} = -\left[(\rho+P)\sinh^2\alpha + p + 2q_1\frac{\sinh\alpha}{R}\right] - \frac{1}{2}f C_4^2, \tag{7}$$

$$\frac{R_4^2}{R^2} + 2\frac{R_{44}}{R} = -p - \frac{1}{2} f C_4^2, \tag{8}$$

$$3\frac{R_4^2}{R^2} + \frac{2\beta R_4}{R^4\left(1+\beta\int\dfrac{dt}{R^3}\right)} = (\rho+P)\cosh^2\alpha - p + 2q_1\frac{\sinh\alpha}{R} + \frac{1}{2} f C_4^2, \tag{9}$$

$$(\rho+P)R\sinh\alpha\cosh\alpha + q_1\cosh\alpha + q_1\frac{\sinh^2\alpha}{\cosh\alpha} = 0. \tag{10}$$

### 3. Tilted Model

We assume that the model is filled with stiff fluid

$$p = \rho. \tag{11}$$

Equation (7), (9) and (11) we get

$$R = mT^n, \tag{12}$$

$$\left(1+\beta\int\frac{dt}{R^3}\right)=\frac{a}{m^3 n T^{3n-1}},\tag{13}$$

where $m=\left(\frac{1}{n}\right)^n$, $n=\frac{a}{3a+\beta}$, and $T=bt+c$, $b=1$. And a, b, c are integration constant.

The metric (1) reduces to the following form

$$ds^2 = dt^2 - m^2 T^{2n}\left\{dx^2 + dy^2 + \left(\frac{a}{m^3 n T^{3n-1}}\right)^2 dz^2\right\}.\tag{14}$$

Where $T=bt+c$, $b=1$, $m=\left(\frac{1}{n}\right)^n$, and $n=\frac{a}{3a+\beta}$.

The conservation equation for c-field given by

$$\left(2f\,C_4\,C_{44}\right)+2\left[\frac{3n}{T}+\frac{n\beta}{aT}\right]f\,C_4^2 = 0.\tag{15}$$

## 4. Physical Properties

Equations (15) we get

$$C = s^{1/2}\log T + l,\tag{16}$$

where $s$ and $l$ are integration constant.

From equation (8), (11) - (13) and (16)

$$p = \rho = \frac{2n(2-3n)-f\,s}{2T^2}.\tag{17}$$

Solving equation (7), (9) - (13),(12),(16) and (17) we get the tilt angle $\alpha$ is given by

$$\cosh\alpha = \left[\frac{3na(1-n)+\beta n^2 - a\,f\,s}{2n(a+\beta n)-a\,f\,s}\right]^{1/2}\tag{18}$$

$$\sinh\alpha = 0.\tag{19}$$

From equation (7), (8), (12), (17) and (18) we get the heat conduction vectors $q_i$ is

$$q^1 = 0,\tag{20}$$

$$q^4 = 0. \tag{21}$$

The flow vectors $u^i$ is given by

$$u^1 = 0, \tag{22}$$

$$u^4 = \left[\frac{3na(1-n) + \beta n^2 - a f s}{2n(a+\beta n) - a f s}\right]^{1/2}. \tag{23}$$

The scalar of expansion, shear scalar and as

$$\theta = \frac{1}{T}\left[\frac{3na(1-n) + \beta n^2 - a f s}{2n(a+\beta n) - a f s}\right]^{1/2}, \tag{24}$$

$$\sigma^2 = \frac{2n^2\beta^2}{3a^2 T^2}. \tag{25}$$

The rate of expansion $H_i$ in the direction of $x$, $y$, $z$-axis are given by

$$H_1 = H_2 = \frac{2n}{T}, \tag{26}$$

$$H_3 = \frac{2n(a+\beta)}{aT}. \tag{27}$$

The density parameters is given by

$$\Omega = \frac{2n(2-3n) - f s}{24}. \tag{28}$$

When T= 0, the creation field C is $-\infty$ and for large value of T, the creation field C is NaN (Not a Number, MATLAB represent that are not a real or complex with special value called Not a Number).initially value of T, the pressure and density are infinite but at $T = \infty$, the pressure and density are vanish .The heat conduction vector $q^1$ and $q^4$ are zero.. Tilt angle $\alpha$, flow of vector $u^1$ and $u^4$ and density parameter $\Omega$ are constant. For large value of T the scalar of expansion and shear scalar are zero but at T = 0 the scalar of expansion and shear scalar are infinite. The rate of expansion $H_i$ in the direction of $x$, $y$, $z$-axis is vanishing at $T = \infty$ but initially the rate of expansion $H_i$ in the direction of $x$, $y$, $z$-axis is infinite.

## 5. Non Tilted Model

From equation (7) and (8)

$$\frac{2q_1}{R} = -(\rho+p)\sinh\alpha . \tag{29}$$

Using equation (29), the Einstein's field equation (7),(9) and (10) reduces to

$$\frac{R_4^2}{R^2} + 2\frac{R_{44}}{R} = -p - \frac{1}{2}f C_4^2 , \tag{30}$$

$$3\frac{R_4^2}{R^2} + \frac{2\beta R_4}{R^4\left(1+\beta\int\frac{dt}{R^3}\right)} = \rho + \frac{1}{2}f C_4^2, \tag{31}$$

$$(\rho+P)R = 0 . \tag{32}$$

$$\rho_4 + \left[3\frac{R_4}{R} + \frac{\beta}{R^3\left(1+\beta\int\frac{dt}{R^3}\right)}\right](\rho+p) = -f C_4\left[C_{44} + \left(\frac{3R_4}{R} + \frac{\beta}{R^3\left(1+\beta\int\frac{dt}{R^3}\right)}\right)C_4\right] , \tag{33}$$

The source of equation of C-field $C_4 = 1$, equation (30)-(32) we get

$$\frac{R_4^2}{R^2} + 2\frac{R_{44}}{R} = -p - \frac{1}{2}f , \tag{34}$$

$$3\frac{R_4^2}{R^2} + \frac{2\beta R_4}{R^4\left(1+\beta\int\frac{dt}{R^3}\right)} = \rho + \frac{1}{2}f , \tag{35}$$

$$(\rho+P)R = 0 . \tag{36}$$

Equation (36) we get three different Universes

### Case-I   Einstein Universes

We consider $R = 0$

Equation (34) and (35) we get

$$p = \rho = -\tfrac{1}{2} f . \tag{37}$$

Equation (33) is given by

$$C_4 = \left(\frac{2h}{f}\right)^{1/2} , \tag{38}$$

Where h is integration constant.

## Case-II  De-sitter Universe

We assume $\rho + p = 0, \quad R \neq 0$

Equation (34) and (35) we get

$$2\frac{R_4^2}{R^2} + \frac{R_{44}}{R} + \frac{\beta R_4}{R^4 \left(1 + \beta \int \frac{dt}{R^3}\right)} = 0 \tag{39}$$

Solving equation (39) we get

$$R = M T^N, \quad \left(1 + \beta \int \frac{dt}{R^3}\right) = \frac{a_1}{M^3 N T^{3N-1}} . \tag{40}$$

where $M = \left(\frac{1}{N}\right)^N$, $N = \frac{a_1}{3a_1 + \beta}$, and $T = b_1 t + c_1$, $b_1 = 1$. And $a_1$, $b_1$, $c_1$ are integration constant.

The metric (1) reduces to the following form

$$ds^2 = dt^2 - M^2 T^{2N}\left\{dx^2 + dy^2 + \left(\frac{a_1}{M^3 N T^{3N-1}}\right)^2 dz^2\right\} . \tag{41}$$

Equation (33) is given by

$$C = s_1^{1/2} \log T + l_1 , \tag{42}$$

where $s_1$ and $l_1$ are integration constant.

The scalar of expansion, shear scalar and as

$$\theta = \frac{1}{T}, \tag{43}$$

$$\sigma^2 = \frac{2N^2 \beta^2}{3a_1^2 T^2}. \tag{44}$$

The rate of expansion $H_i$ in the direction of $x$, $y$, $z$-axis are given by

$$H_1 = H_2 = \frac{2N}{T}, \quad H_3 = \frac{2N(a_1 + \beta)}{a_1 T}. \tag{45}$$

When T= 0, the creation field C is $-\infty$ and for large value of T, the creation field C is NaN. For large value of T the scalar of expansion and shear scalar are zero but at T = 0 the scalar of expansion and shear scalar are infinite. The rate of expansion $H_i$ in the direction of $x$, $y$, $z$-axis is vanishing at $T = \infty$ but initially the rate of expansion $H_i$ in the direction of $x$, $y$, $z$-axis is infinite.

### Case-III

We assume $\rho + p = 0$, $R = 0$

Equation (33) is given by

$$C = \left(\frac{2k}{f}\right)^{1/2}, \tag{46}$$

where k is integration constant.

### Conclusion

The study results into an expanding and shearing universe. It observed that the universe starts with big-bang at T = 0 and decreases with the age of the universe increases. For tilted models initially, the pressure, density, the scalar of expansion and shear scalar are infinite. When $T = \infty$ the pressure, density, the scalar of expansion and shear scalar are vanish. Tilt angle $\alpha$, flow of vector $u^4$ and density parameter $\Omega$ are constant. The heat conduction vector $q^1, q^4$ and the flow of vector $u^1$ are zero. The rate of expansion $H_i$ in the direction of $x$, $y$, $z$-axis is vanishing at $T = \infty$ but initially the rate of expansion $H_i$ in the direction of *x, y, z*-axis is infinite. When T= 0, the creation field C is $-\infty$ and for large value of T, the creation field C is NaN (Not a Number, MATLAB represent that are not a real or complex with special value called Not a Number). since $\lim_{T \to \infty}\left(\frac{\sigma}{\theta}\right) \neq 0$ the models not approach isotropy for large value of T. At n = 0, the shear scalar are zero. For n = 0, $\lim_{T \to \infty}\left(\frac{\sigma}{\theta}\right) = 0$ the models approach isotropy for large value of T.

**For non tilted models**:

In case-I, the pressure and density are constant.

In case-II, The study results into an expanding and shearing universe. It observed that the universe starts with big-bang at T = 0 and decreases with the age of the universe increases. When T= 0, the creation field C is $-\infty$ and for large value of T, the creation field C is NaN. For large value of T the scalar of expansion and shear scalar are zero but at T = 0 the scalar of expansion and shear scalar are infinite. The rate of expansion $H_i$ in the direction of *x, y, z*-axis is vanishing at $T = \infty$ but initially the rate of expansion $H_i$ in the direction of *x, y, z*-axis is infinite. since $\lim_{T \to \infty}\left(\frac{\sigma}{\theta}\right) \neq 0$ the models not approach isotropy for large value of *T*. At N = 0, the shear scalar are zero. For N = 0, $\lim_{T \to \infty}\left(\frac{\sigma}{\theta}\right) = 0$ the models approach isotropy for large value of *T*.

In case-III, creation field C is constant.

**References**

[1] H.Bondi, T. Gold, (1948) Mon. Not. R. Astron. Soc. 108, 252 .